\documentclass[pre,aps,floats,superscriptaddress,floatfix,twocolumn]{revtex4}
\usepackage{amssymb,amsmath}
\usepackage{amsmath,amssymb}
\usepackage{graphicx}
\usepackage{psfrag}
\usepackage{color}
\usepackage{dcolumn}
\usepackage[normalem]{ulem}

\def\beq{\begin{equation}}
\def\eeq{\end{equation}}
\def\bea{\begin{eqnarray}}
\def\eea{\end{eqnarray}}
\begin{document}
\title{Universal scaling regimes in rotating fluid turbulence}
 \author{Abhik Basu}\email{abhik.basu@saha.ac.in,abhik.123@gmail.com}
\affiliation{Theory Division, Saha Institute of
Nuclear Physics, Calcutta 700064, West Bengal, India}
\author{Jayanta K Bhattacharjee}\email{jayanta.bhattacharjee@gmail.com}
\affiliation{Department of Theoretical Physics, Indian Association for the Cultivation of Science, 2A and 2B Raja S C Mullick 
Road, Calcutta 700032, West Bengal, India}
\date{\today}
\begin{abstract}
We analyse the scaling properties of the energy spectra in fully developed incompressible
 turbulence in forced, rotating
fluids in three dimensions (3D), which are believed to be 
characterised 
by 
{\em universal 
scaling exponents} in the inertial range. To elucidate the scaling regimes, we set up a scaling analysis of the 3D Navier-Stokes equation for a rotating fluid that is driven by large-scale external forces. We use scaling arguments to extract the scaling exponents, which characterise the different scaling regimes of the energy spectra. We speculate on the intriguing possibility of two-dimensionalisation of 3D rotating turbulence within our scaling theory. Our results can be tested in large scale simulations and 
relevant laboratory-based experiments.

\end{abstract}

  \maketitle
  
  \section{Introduction}
  
  Nonequilibrium
systems are described by the appropriate equations of motion for the relevant dynamical variables and
exhibit much richer universal behavior than usually observed in equilibrium critical dynamics~\cite{hal}.
Hydrodynamic turbulence in fluids, described by the Navier-Stokes
equation~\cite{land,frish} for the evolution of the velocity field
$\bf v$, is a prime example of an out of equilibrium system, due to
the external drive acting on the fluids. Interestingly, fully developed fluid
turbulence in three- (3D) and two- (2D) dimensions show markedly
different behavior: In 3D, the energy spectra follow the well-known K41 result for homogeneous and isotropic
3D hydrodynamics turbulence where the one-dimensional energy spectra $E(k)\sim k^{-5/3}$ (hereafter K41) in the inertial range, where $k$ is a wavevector~\cite{k41}. This K41 result is quite robust and universal, and found in wide-ranging natural systems, e.g., shear flows~\cite{shear}, viscoelastic fluids~\cite{visco} and jet flows~\cite{jet}. In contrast, 2D turbulence is characterised by an inverse cascade of energy at very large scales with $E(k)\sim k^{-5/3}$, and forward cascade of enstrophy with $E(k)\sim k^{-3}$ at intermediate scales~\cite{2d1,2d2,2d3,2d4,2d5}. 

Rotating turbulence, i.e., turbulence in a rotating fluid, is a naturally occurring phenomenon in many astrophysical and geophysical flows, as well as in laboratory-based engineering fluid flows. The presence of the Coriolis forces is the distinctive feature of rotating turbulence, which should affect the large-scale scaling properties of rotating turbulence. In spite of extensive studies, there is still no good agreement on the scaling of the energy spectra in rotating turbulence, in particular how the Coriolis forces affect the scaling of the spectra at very large scales.  
 Varieties of analytical, numerical, or experimental investigations of
either the forced or the decaying rotating turbulent fluid suggest that the kinetic energy spectra in the rotation-dominated small-$k$ regions should scale as $E(k)\sim k^{-m},\,m\in (2,3)$~\cite{2ds1,2ds2,2ds3,2ds4,2ds5,2ds6,2ds7,2ds8,2ds9,2ds10,2ds11,2ds12,2ds13,2ds14,2ds15}. Recent perturbative studies indicate that the one-dimensional kinetic energy spectra made out of the velocity component parallel to the rotation axis scales as $k^{-5/3}$, indistinguishable from the K41 prediction. In contrast, the one-dimensional kinetic energy spectra made out of the velocity components lying in a plane normal to the rotation axis scales as $k^{-3}$, different from the K41 scaling~\cite{ab-jkb-rot}.  The precise forms of the scaling of the energy spectra in rotating turbulence however is still not well-settled.

There is a degree of formal similarity between the (linearised) equations of motion of rotating turbulence, which is nothing but the Navier-Stokes equation in a rotating frame (see below) and the equations for magnetohydrodynamic turbulence (MHD) in the presence of a mean magnetic field $B_0$~\cite{jackson,arnab}. In the former case, the Coriolis forces lead to oscillatory modes, whereas in the MHD case, a non-zero $B_0$ gives rise to propagating Alf\'ven waves. Strong Alf\'ven waves are known to make the energy spectra in MHD anisotropic, and change the scaling as well~\cite{jkb-mhd}. In the same vein, strong Coriolis forces should make the scaling of energy spectra in rotating turbulence anisotropic and also {\em different} from its isotropic counterpart (i.e., the K41 scaling).

In this work, we revisit the universal scaling of energy spectra in forced, statistically steady rotating turbulence in its inertial range. To this end, we have set up a scaling theory to study the scaling of the energy spectra in the inertial range. We cover both the weak and strong rotation limits. In the former case, unsurprisingly, the K41 result is found. With stronger rotation, anisotropic scaling with different exponents ensues. In particular, in the wavevector region $k_\perp \gg k_\parallel$, our scaling theory gives the scaling of the 2D spectra $E(k_\perp,k_\parallel)$, where ${\bf k}_\perp$ and $k_\parallel$ are the components of the wavevector $\bf k$ in the plane perpendicular to the rotation axis (here $\hat z$-axis) and along the rotation axis, respectively. We find $E(k_\perp,k_\parallel)\sim k_\perp^{-5/2}k_\parallel^{-1/2}$ for $k_\perp \gg k_\parallel$, which agrees with the Kuznetsov–Zakharov–Kolmogorov spectra predicted by the weak inertial-wave turbulence theory for the rotating fluids~\cite{2ds7}.
We show that this result is unaffected by nonlinear fluctuation corrections at the one-loop order. We further demonstrate that this result could be obtained by demanding that the cascade of the kinetic energy flux is hindered by a non-zero helicity, which is naturally present in a rotating fluid. In the opposite limit of $k_\perp\ll k_\parallel$, we get $E(k_\perp,k_\parallel)\sim k_\perp^{-1}k_\parallel^{-2}$. We also show perturbatively in the rotation $\Omega$ that the kinetic energy flux is indeed reduced by it. The remainder of the article is organised as follows. In Sec.~\ref{rot-ns}, we set up the forced Navier-Stokes equation in a rotating fluid. Then in Sec.~\ref{scale} we set up the scaling arguments. Then next in Sec.~\ref{iso-scale}, we revisit the K41 scaling scaling in an isotropic, nonrotating fluid turbulence, and show how are scaling theory reproduces it. Next, in Sec.~\ref{weak-scale}, we show that for weak rotation, the energy spectra again show the K41 scaling. Then in Sec.~\ref{strong-scale} we study the 2D anisotropic energy spectra in the opposite limit of large $\Omega$. In Sec.~\ref{summ} we discuss and summarise our results. We provide some technical results, including a perturbative demostration of the reduction of the kinetic energy flux by helicity, in the Appendix for interested readers.
  
  
  \section{Turbulence in a rotating fluid}\label{rot-ns}
  
\begin{figure}[htb]
 \includegraphics[width=\columnwidth]{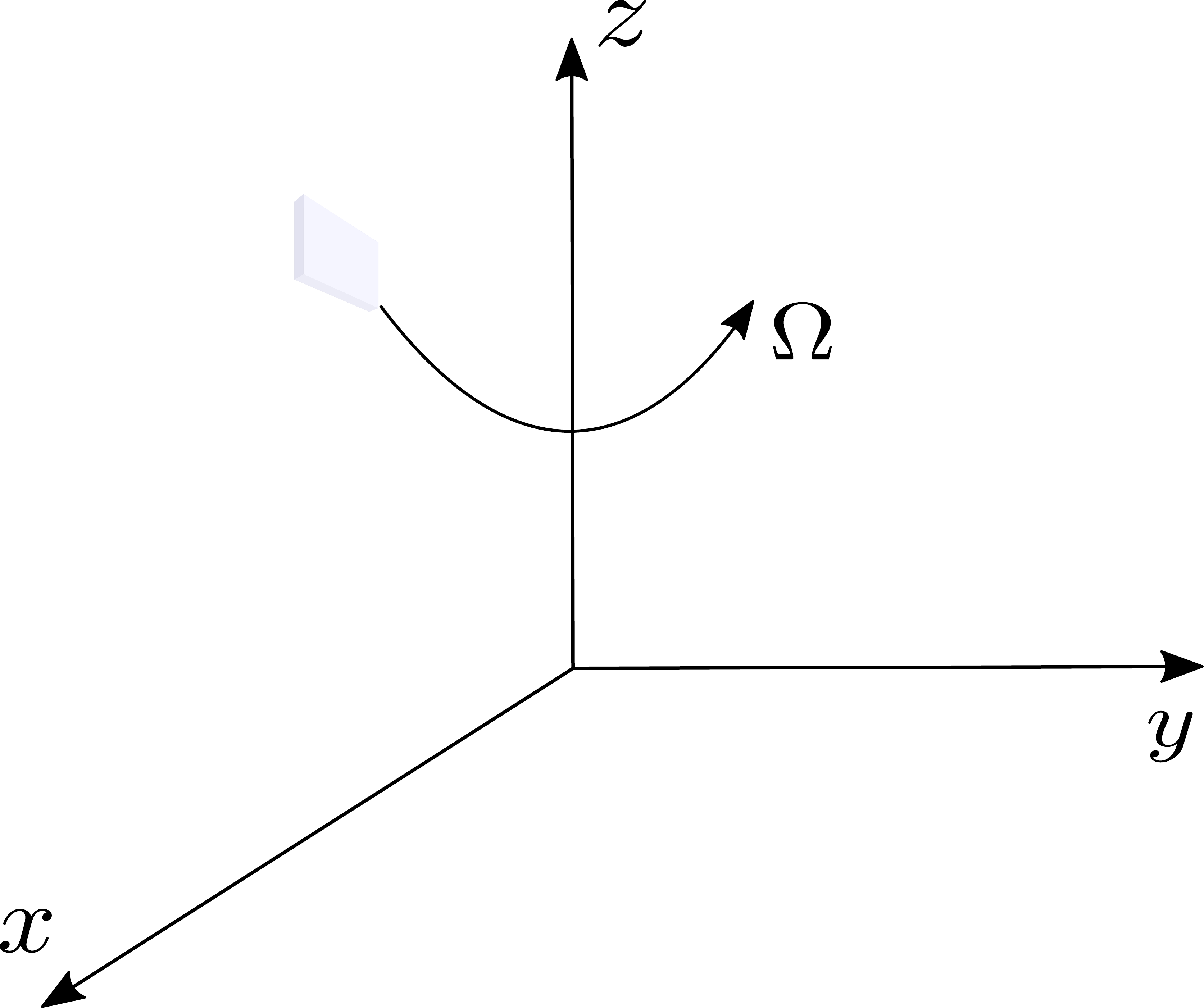}
 \caption{Geometry of the rotating fluid. We assume the rotation to be about the $z$-axis. In this coordinate system, ${\bf k}_\perp = (k_x,k_y)$ and $k_\parallel = k_z$.}\label{rot-pic}
\end{figure}

  The Navier-Stokes equation for the velocity field ${\bf v} ({\bf r},t)$ in a rotating frame with rotation ${\boldsymbol \Omega} = \omega \hat z$ is given by
    \begin{equation}
   \frac{\partial\bf v}{\partial t} +2 ({\boldsymbol\Omega}\times {\bf v}) + \lambda ({\bf v}\cdot {\boldsymbol\nabla}){\bf v} = - \frac{{\boldsymbol\nabla}p^*}{\rho} + \nu\nabla^2 {\bf v} + {\bf f}, \label{rot-NS}
  \end{equation}
where $p^*=p+ \frac{1}{2}|{\boldsymbol\Omega}\times {\bf v}|^2$ is the effective pressure. We assume ${\boldsymbol\Omega}=\Omega \hat z$, i.e., the rotation is about the $z$-axis; see Fig.~\ref{rot-pic}. 

In this case (\ref{rot-NS}) may be written in terms of components as
\begin{eqnarray}
 &&\frac{\partial v_z}{\partial t} + \lambda({\bf v}\cdot {\boldsymbol\nabla})v_z = -\frac{\partial_z p^*}{\rho}+\nu \nabla^2 v_z + f_z, \label{vz-eq}\\
 &&\frac{\partial v_x}{\partial t} - 2\Omega v_y + \lambda({\bf v}\cdot {\boldsymbol\nabla})v_x = -\frac{\partial_x p^*}{\rho} + \nu\nabla^2 v_x + f_x,\label{vx-eq}\\
  &&\frac{\partial v_y}{\partial t}  +2\Omega v_x + \lambda({\bf v}\cdot {\boldsymbol\nabla})v_y = -\frac{\partial_y p^*}{\rho} + \nu\nabla^2 v_y + f_y \label{vy-eq}.
\end{eqnarray}
Here, $\lambda=1$.
These equations in 3D admit two conserved quantities in the inviscid limit: (i) kinetic energy  $E=\int d^3x \rho\, v^2/2$ and (ii) helicity $H= \int d^3x\, {\bf v}\cdot {\boldsymbol\nabla}\times {\bf v}={\bf v}\cdot {\boldsymbol\omega}$, where $\boldsymbol\omega\equiv {\boldsymbol\nabla}\times {\bf v}$ is the local vorticity.  In the viscous steady states, in a Kolmogorov-like picture neglecting intermittency, $E$ and $H$ should have constant (i.e., scale independent) fluxes. Clearly, $E/H$ has the dimension of a length, which allows us to define a length-scale $l^*=E/H$. We assume the external forces to be non-helical, i.e., no helicity injection by the forces. Thus helicity is generated in the bulk only by the global rotation.  We consider the incompressible limit, i.e., the mass density $\rho=const.$, or equivalently ${\boldsymbol\nabla}\cdot {\bf v}=0$.  At this stage, it is useful to set up the notations. Below we use $\tilde \omega$ and $\tilde \Omega$ to denote Fourier frequencies, while $\Omega$ and $\boldsymbol\omega$ represent the global rotation frequency and vorticity, respectively.
  
  \section{Scaling analysis} \label{scale}
  
  
  To classify the scaling regimes, we first define the following dimensionless numbers 
  
  (i) Rossby number $R_o=U/(2\Omega L)$, 
  
  (ii) Reynolds number $R_e=\frac{LU}{\nu}$ and
  
  (iii) Ekman number $Ek=R_o/R_e=\nu/(2\Omega L^2)$,
  
  where $L$ is the linear system size, and $U$ is a typical velocity. 
  We expect to find two distinct scaling regimes as characterised by $R_o$ (or $\Omega$): 
  
  (i) Weak rotation $\Omega \rightarrow 0$, or $R_o\rightarrow\infty$


  (iii) Large rotation $\Omega \rightarrow \infty$, or $R_o\rightarrow 0$.
  
  Since the rotation picks up a direction (the axis of rotation, here the $z$-axis), system is generally anisotropic. We therefore construct an anisotropic scaling theory of the system: we assume scaling under the transformations
  \begin{equation}
   {\bf r}_\perp \rightarrow l_\perp {\bf r}_\perp,\,z\rightarrow l_\perp^\xi z,\, {\bf v}_\perp \rightarrow l_\perp^{a_\perp} {\bf v}_\perp,\, v_z \rightarrow l_\perp ^{a_z}v_z.\label{basic-scaling}
  \end{equation}
  Here, ${\bf r}_\perp \equiv (x,\,y)$, ${\bf v}_\perp \equiv (v_x,\,v_y)$.
In a general anisotropic situation,  $\xi\neq 1$. We also allow for the possibility $a_\perp \neq a_z$, i.e., ${\bf v}_\perp$ and $v_z$ may not scale in the same way under spatial rescaling. We further define time-scale $t$ to scale as
\begin{equation}
 t\sim l_\perp^{\tilde z},
\end{equation}
where $\tilde z$ is a dynamic exponent. Furthermore, we define a phenomenological dimensionless constant $\tilde \Omega$ by
\begin{equation}
 \tilde \Omega = \frac{[{\boldsymbol\Omega}]}{[ {\boldsymbol\omega}]},
\end{equation}
Here, [...] implies ``in a dimensional sense''~\cite{jkb-mhd}. Clearly, the two limiting cases $\tilde \Omega\rightarrow 0$ and $\tilde \Omega \rightarrow \infty$ phenomenologically correspond to $R_o\rightarrow \infty$ and $R_o\rightarrow 0$. 

We note that by balancing the Coriolis force terms against the advective nonlinear terms, we can extract a length-scale $L_o$. In a scaling sense, we set
\begin{equation}
 \Omega v \sim \frac{v^2}{L_0},
 \end{equation}
giving $v \sim \Omega L_0$. Dimensionally speaking, energy dissipation
\begin{equation}
 \epsilon\sim \frac{v^3}{L_0}\sim \Omega^3 L_0^2.
\end{equation}
This gives
\begin{equation}
 L_0\sim \sqrt{\frac{\epsilon}{\Omega^3}}.
\end{equation}
The corresponding wavevector $k_0\equiv 2\pi/L_0$ is the {\em Zeeman wavevector}. 
For length-scales $L\gg L_0$ (or equivalently, for wavevector $k\ll k_0$), we expect Coriolis force terms to be important, effects of the rotation should be strong, and hence non-K41 spectra should follow. In the opposite limit of $L\ll L_0$ (or $k\gg k_0$), Coriolis forces should be irrelevant, and hence K41 scaling should follow. Thus a dual scaling is believed to exist~\cite{2ds12,dual1,dual2,dual3}. { In terms of the dimensionless numbers, we are interested in $R_e\rightarrow \infty$ for fully developed turbulence. Together with $R_e\rightarrow \infty$ (implying fully developed turbulence) and $Ek\rightarrow 0$ (implying Coriolis forces dominating over the viscous damping at large scales), we can have two situations: (i) $R_o\rightarrow \infty$ for weak rotation, and (ii) $R_o\rightarrow 0$ for strong rotation. Lastly, one has the dissipation scale $\eta_d$, such that for length scales smaller than $\eta_d$, the dissipation range ensues. Then in terms of the length scales defined above, we can have the following scenarios.}
In a sufficiently large system, there should be adequate scale separations, such that $\eta_d  \ll L_0$, i.e., $L_0$ should belong to the inertial range. This should allow for both the scaling regimes, {\em viz.} K41 and non-Kolmogorov scaling regimes to be observed.  


\subsection{Nonrotating isotropic case}\label{iso-scale}

For a nonrotating, isotropic fluid, $\Omega=0$ in (\ref{vx-eq})-(\ref{vy-eq}) gives the usual 3D isotropic Navier-Stokes equation. Let us briefly revisit the extraction of the K41 scaling by applying the scaling arguments on the usual 3D Navier-Stokes equation first. Due to the isotropy of the system, we expect $\chi=1$ strictly, and make no distinction between $l_\perp$ and $l_\parallel$, the rescaling factors of ${\bf r}_\perp$ and $z$ respectively: $l_\perp \sim l_\parallel \sim l$. Demanding scale invariance~\cite{jkb-mhd}, we find
\begin{equation}
 \frac{\partial {\bf v}}{\partial t}\sim {\bf v}\cdot {\boldsymbol\nabla} {\bf v} \implies l^{a-\tilde z}= l^{2a-1} \implies a=1-\tilde z.
\end{equation}
Next, in a mean-field like approach, we assume the kinetic energy flux or the kinetic energy dissipation per unit mass is scale invariant in the inertial range. This gives
\begin{equation}
 \frac{\partial v^2}{\partial t}\sim l^0 \implies 2a=\tilde z.
\end{equation}
Combining then, we get $a=1/3,\,\tilde z=2/3$. This corresponds to a 1D kinetic energy spectra $E(k)\sim k^{-5/3}$, the expected K41 result. 
  
  \subsection{Weak rotation effects: $k\gg 2\pi/L_0$}\label{weak-scale}
  
  In order to study the effects of weak rotation on the scaling of the energy spectra, we consider the limit $R_o\rightarrow \infty$, or $ \Omega \rightarrow 0$. Equivalently, we consider length scales $L\ll L_0$, with the understanding that $L\gg \eta_r$, the dissipation scale. In this case, the Coriolis force is unimportant.  Hence $\langle H\rangle \approx 0$, where $<...>$ implies averages over the statistical steady states. Thus, the flux of $E$ is the relevant (in the Kolmogorov sense) flux. Then, proceeding as in Ref.~\cite{jkb-mhd}, we unsurprisingly recover the K41 scaling:
  \begin{equation}
   a_\perp = a_z = 1/3,\,\tilde z=2/3,\,\xi=1.
  \end{equation}
The last of the above naturally means isotropic scaling (although geometry remains anisotropic). Furthermore, if we let $R_o\sim l^\eta$ and demand scale invariance of all the terms (including the Coriolis force terms), we find
\begin{equation}
 \eta = \tilde z = 2/3.
\end{equation}
Thus $R_o\rightarrow \infty$ as $l\rightarrow \infty$, which corresponds to nonrotating and isotropic fully developed turbulence. We have assumed that the nonlinear coupling constant does not scale under spatial rescaling, which is consistent with the nonrenormalisation of $\lambda$ due to the Galilean invariance of the Navier-Stokes equation. The scaling of the viscosity is controlled by the dynamic exponent $\tilde z$. 
  
  \subsection{Strong rotation effects: $k\ll 2\pi/L_0$}\label{strong-scale}
  
  We next consider the large rotation case, i.e., when $R_o\geq {\cal O}(1)$, or $ \Omega \geq  {\cal O}(1)$. It is now expected that anisotropy is significant. Below we analyse the scaling in several equivalent ways. 
  
  Balancing different terms of (\ref{vx-eq}) and (\ref{vy-eq}) we find
  \begin{equation}
   \tilde z=0,\,a_\perp =1,\,a_z=\xi.\label{expo-list1}
  \end{equation}
  To proceed further, we allow for the possibility that not only the spatial scaling may be anisotropic, there may be different dynamic exponents for ${\bf v}_\perp$ and $v_z$, with $\tilde z$ being identified as the dynamic exponent $\tilde z_\perp$ of ${\bf v}_\perp$. To study this, we separately consider the contribution to the kinetic energy from the in-plane velocity ${\bf v}_\perp$ and normal component of the velocity $v_z$.
  Interestingly, the exponents in (\ref{expo-list1}) mean that the flux of the ``in-plane kinetic energy'' $E_\perp\equiv \int d^3 x\, v_\perp^2$ cannot be scale-independent! Let us now consider the kinetic energy $E_z\equiv \int d^3x\, v_z^2$ flux of $v_z$. If we assume $\tilde z=0$ is the dynamic exponent of $v_z$ also, then $a_z=0=\xi$ can actually keep the flux of $E_z$ scale-independent. However, $a_z=0$ is unexpected, as it means $v_z$ {\em does not} scale with $l$ at all. Assume the dynamics $z_\parallel$ of $v_z$ be non-zero: $z_\parallel >0$. Now consider Eq.~(\ref{vz-eq}) and balance 
  \begin{equation}
   \frac{\partial v_z}{\partial t} \sim v_z\partial_z v_z \implies a_z=1-z_\parallel.
  \end{equation}
Next, demanding scale-invariance of the flux of $E_z$ gives 
\begin{equation}
 2a_z=\tilde z_\parallel \implies a_z=\frac{1}{3},\,\tilde z_\parallel = \frac{2}{3}.
\end{equation}
This further means $\xi=a_z=1/3$. Notice that with $a_\perp =1,\,\xi=1/3$, $ ({\bf v}_\perp\cdot {\boldsymbol\nabla}_\perp) v_z$ and $v_z\partial_z v_z$ scale the same way. Since we get $a_\perp >a_z$, we should have $\langle v_\perp^2\rangle \gg \langle v_z^2\rangle$ in the long wavelength limit, suggesting concentration of the kinetic energy in a plane normal to the rotation axis~\cite{gode}. This implies an effective {\em two-dimensionalisation}.  On the other hand, $\tilde z<\tilde z_\parallel$ implies ${\bf v}_\perp \ll v_z$ in the long time limit, a conclusion contradictory to our above inference. In fact, this alternative scenario implies a type of dimensional reduction, where most of the energy is confined to the $z$-direction. We are unable to conclusively predict which of these two scenarios actually holds. Numerical studies should be useful in this regard.

Is the ensuing flow field in the limit $R_o\rightarrow \infty$ truly 2D? In our opinion, the answer is no. First of all, the flow remains overall 3D incompressible. This means the effective 2D flow field might be {\em 2D compressible}, which is an interesting possibility. Secondly, it is not whether the direction of the kinetic energy cascade becomes backward, a hallmark of pure 2D turbulence. Thirdly, enstrophy is a conserved quantity in the inviscid limit of pure 2D turbulence, whereas it is not expected to be so in the 3D rotating case even in the limit of high rotation. Therefore, notwithstanding the dominance of $v_\perp$ over $v_z$, the resulting flow field should be fundamentally different from pure 2D nonrotating turbulence.  
Lastly, since $z_\perp\neq z_\parallel$, we find weak dynamic scaling~\cite{jkb-mhd,dib}. We now calculate the scaling of the two-dimensional kinetic energy spectra $E_\perp(k_\perp,k_\parallel)$ and $E_z(k_\perp,k_z)$, such that total kinetic energy $E_\text{tot}=\int dk_\perp dk_z [E_\perp(k_\perp,k_z)+E_z(k_\perp,k_z)]$. The scaling of $E_\perp(k_\perp,k_z)$ and $E_z(k_\perp,k_z)$ can be obtained as follows. We use the general definition to write in a dimensional/scaling sense
\begin{eqnarray}
 &&v_m({\bf k},\tilde\omega)\sim \int v_m({\bf x},t)\exp(i{\bf k_\perp}\cdot {\bf x_\perp})\nonumber \\&&\times\exp (ik_\parallel z) \exp(i\tilde\omega t)d^2 x_\perp dz dt \nonumber \\&\sim& l_\perp^{a_m}l_\perp^2 l_\parallel\sim \frac{1}{k_\perp^{a_m+2}}\frac{1}{k_\parallel}.
\end{eqnarray}
Here, $m=\perp,\,\parallel$. Furthermore,
\begin{eqnarray}
 &&\langle {\bf v}_\perp ({\bf k}_1)\cdot {\bf v}_\perp ({\bf k_2})\rangle = F_\perp(k_1)\delta({\bf k}_1+{\bf k}_2),\\
 &&\langle { v}_z ({\bf k}_1) { v}_z ({\bf k_2})\rangle = F_\parallel(k_1)\delta({\bf k}_1+{\bf k}_2).
\end{eqnarray}
Dimensionally then,
\begin{equation}
 F_\perp\sim \frac{1}{k_\perp^{a_\perp+2}k_\parallel},\,F_\parallel\sim \frac{1}{k_\perp^{a_\parallel+2}k_\parallel}.
\end{equation}
This gives for the two-dimensional energy spectra
%
%
\begin{equation}
 E_\perp(k_\perp,k_\parallel)\sim k_\perp^{-3},\,E_z(k_\perp,k_z)\sim k_\perp^{-5/3}. \label{spec2d}
\end{equation}
If we ignore anisotropy, we can define two corresponding one dimensional energy spectra $E_\perp(k)$ and $E_z(k)$ from (\ref{spec2d}) by $E_\text{tot} = \int dk [E_\perp(k) + E_z(k)]$.
Notice that, neglecting anisotropy, the one-dimensional spectra corresponding to $ E_\perp(k)\sim k_\perp^{-3}$ should scale as $k^{-2}$ as argued above, in agreement with Refs.~\cite{dual1,dual2,dual3}. Nonetheless, in spite of this agreement, we notice that our results (\ref{spec2d}) appear to suggest that $E_\perp(k_\perp,k_z)$ and $E_z(k_\perp,k_z)$ have {\em no} $k_z$-dependence, which should be unphysical.  We try to rectify this below.

  
  First of all, for large $\Omega$, the scaling should be dominated by the Coriolis forces. The vorticity ${\boldsymbol\omega}$ satisfies~\cite{2ds7} 
  \begin{equation}
   \partial_t {\boldsymbol\omega}({\bf k})=-2\Omega\frac{k_\parallel \hat e_{\bf k}\times {\boldsymbol\omega}_{\bf k}}{k},
  \end{equation}
  giving time-scale $\tau\sim k/k_\parallel \sim k_\perp/k_\parallel$ for $k_\perp\gg k_\parallel$.

 It is thus reasonable to assume that $\tau$ as defined above is the relevant time-scale when $R_o\rightarrow 0$. In what follows below, we do not make any distinction between $v_\perp$ and $v_z$. We now impose the scale-independence of the kinetic energy flux $\Pi$. The energy flux may be calculated from the Navier-Stokes equations (\ref{rot-NS}). We find
 
 \begin{widetext}
 \begin{eqnarray}
  \Pi &=& -2\lambda^2\int_{{\bf k,q},\tilde\omega,\tilde \Omega}\bigg[M_{imn}({\bf k})M_{ijp}(-{\bf k})\langle v_m({-k+q}, -\tilde\omega+\tilde\Omega)v_p ({\bf k-q},\tilde\omega-\tilde\Omega)\rangle \\ &+& M_{jmn}({\bf k})M_{ijp}({\bf q}) \langle v_i ({-k},-\tilde\omega)v_n({\bf k},\tilde\omega)\rangle \langle v_m({\bf -k+q},-\tilde\omega+\tilde\Omega)v_p({\bf k-q}, \tilde\omega-\tilde\Omega)\rangle \\&+& 
  M_{jmn}({\bf k-q}) M_{ijp}({\bf k})\langle v_i({-\bf k},-\tilde\omega)v_n({\bf k},\tilde\omega)\rangle \langle v_j({\bf q},\tilde\Omega)v_m({\bf -q},-\tilde\Omega)\rangle\bigg].\label{flux-full}
 \end{eqnarray}
 \end{widetext}
 Here, $M_{ijp}({\bf k})=P_{ij}({\bf k})k_p +P_{ip}({\bf k})k_j$. 
 Since we are interested in the scaling, it suffices to consider the scaling of $\Pi$, suppressing indices and wavevector labels.
 At this one-loop order, suppressing indices and wavevector labels, and assuming $k_\perp \gg k_\parallel$
 \begin{equation}
  \Pi\sim \int d^2 k_\perp d^2p_\perp dk_\parallel dp_\parallel \int dt\int d^2k_\perp\, k_\perp^2 C^2 \sim const., 
 \end{equation}
where $C\equiv \langle v\,v\rangle$ is the correlation function, again suppressing indices and wavevector labels. This gives
\begin{equation}
 C\sim k_\perp^{-7/2}k_\parallel^{-1/2},
\end{equation}
where we have used $\int dt \sim \tau$.
This implies for the two-dimensional kinetic energy spectra
\begin{equation}
 E_\perp(k_\perp,k_\parallel)\sim Ck_\perp \sim k_\perp^{-5/2} k_\parallel^{-1/2},
\end{equation}
as in the wave turbulence theory. It now behooves us to show that the scaling of $\tau$ with wavevector does not get renormalised at the one-loop order. We restrict ourselves here to a scaling-level demonstration.  As shown in Appendix, the one-loop self-energy $\Sigma_{ik}({\bf k},\tilde\omega)$ has the form 
\begin{eqnarray}
 &&\Sigma_{ij}({\bf k},\tilde\omega)\sim \int d^3q d\tilde\Omega M_{imn}({\bf k}) \langle v_m({\bf q},\tilde\Omega) v_r({-\bf q},-\tilde\Omega) \rangle \nonumber \\ &\times&G_{ns}({\bf k-q},\tilde\omega-\tilde\Omega) M_{srj}({\bf k-q}).
\end{eqnarray}
Here, $G_{ns}({\bf k},\tilde\omega)$ is the propagator defined via
\begin{equation}
 G_{ns}({\bf k},\tilde\omega) \equiv \bigg\langle \frac{\delta v_n({\bf k},\tilde\omega)}{\delta f_s({\bf k},\tilde\omega)}\bigg\rangle.
\end{equation}

Considering the one-loop self-energy $\Sigma$ and suppressing indices and wavevector labels [see also Appendix], we obtain
\begin{equation}
 \Sigma\sim k_\perp^2 \int d^2 q_\perp dq_z \int dt G\,C,
\end{equation}
where $G$ is a propagator. Assuming the dominant time-scale in the above time-integral is given by $\tau$, we get (in a scaling sense)
\begin{equation}
 k_\perp^2 \int d^2 q_\perp dq_z \frac{q_\perp}{q_z}q_\perp^{-7/2}k_z^{-1/2}\sim  [k_\perp]^{3/2} [k_\parallel]^{1/2},
\end{equation}
which is less singular than the the bare form of $\tau$. Thus our scaling results on the 2D kinetic energy spectra remain unaffected by the advective nonlinearities in the asymptotic long wavelength limit.

  Interestingly, we can also derive the above results by using phenomenological arguments of suppression of the kinetic energy flux by the helicity generated by the rotation, which are similar to the arguments set up in Ref.~\cite{jkb-mhd} for scaling of the energy spectra in the presence of a strong mean magnetic field in magnetohydrodynamic turbulence. It is known that the predominant role of a (large) non-zero helicity flux is to hinder the cascade of the kinetic energy flux~\cite{kraich}. In fact, it is easy to see from (\ref{vx-eq}) and (\ref{vy-eq}) that a large $\Omega$ should suppress  the nonlinear effects, relative to the Coriolis force terms. Since the nonlinear terms are responsible for the cascade phenomena, we expect the flux of $E_\perp$ to be suppressed by a large $\Omega$. This is similar to the suppression of the energy flux by a strong mean magnetic field in fully developed magnetohydrodynamic turbulence~\cite{jkb-mhd}; see also similar treatment for turbulence in a stably stratified fluid~\cite{ab-jkb-bolgiano}.   
   We write
  \begin{equation}
   \bigg[\frac{\partial v_\perp^2}{\partial t}\bigg]\bigg[\frac{\omega^2}{\Omega^2}\bigg]\sim l^0
  \end{equation}
as the condition of the flux being scale independent. Since dimensionally, $[\omega^2]\sim [v_\perp^2/l^2]$, we get
\begin{equation}
 v_\perp \sim l_\perp^{3/4}l_\parallel^{-1/4}.
\end{equation}
This gives
\begin{equation}
 E_\perp(k_\perp,k_\parallel)\sim k_\perp^{-5/2}k_\parallel^{-1/2},
\end{equation}
in agreement with the conclusion from the wave turbulence theory approaches~\cite{2ds7}. It is easy to get the spectra in the opposite limit $k_\parallel \gg k_\perp$. In this limit, $\tau \sim k_\perp^0 k_\parallel^0$. 
At this one-loop order, suppressing indices and wavevector labels, and assuming $k_\perp \ll k_z$.
 \begin{equation}
  \Pi\sim \int d^2 k_\perp d^2p_\perp dk_z dp_\parallel \int d^2k_\perp\, k_z^2 C^2 \sim const., 
 \end{equation}
Proceeding as before, we find
\begin{equation}
 E_\perp(k_\perp,k_\parallel)\sim k_\perp^{-1}k_\parallel^{-2}.
\end{equation}

  In each of these cases, the corresponding one-dimensional spectra, without making any distinction between $k_\perp$ and $k_\parallel$ scale as $k^{-2}$, consistent with the recent shell-model studies on rotating turbulence~\cite{new-shell}.
  
\begin{figure}[htb]
   \includegraphics[width=\columnwidth]{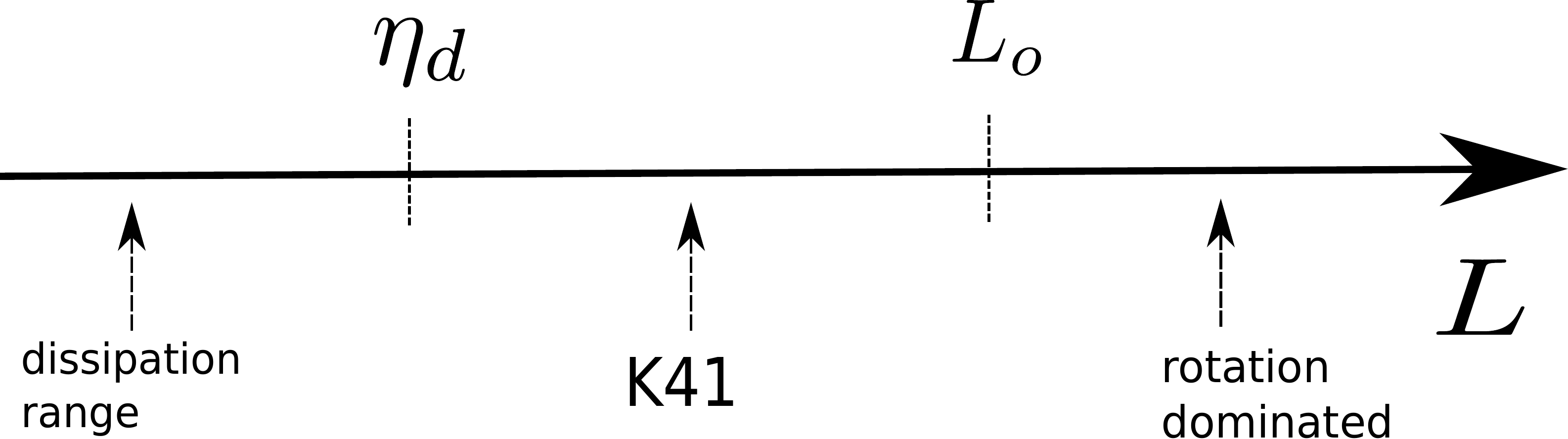}
   \caption{Schematic diagram illustrating the scaling regimes at different length scales. Dissipation range for small scales, K41 scaling regime at the intermediate scales and rotation dominated scaling regimes at the largest scales are shown.}\label{scaling-pic}
  \end{figure}

  A pictorial summary of the scaling regimes are shown in Fig.~\ref{scaling-pic}.
  
  In the above, we have implicitly assumed that the kinetic energy flux $E$ is the relevant flux (in the Kolmogorov sense). As we have discussed above this holds for length scales $\gg l^*$. An interesting case may arise if $l^*\gg L_0$, in which case in the window between $L_0$ and $l^*$, helicity flux $H$ dominates over $E$. If this  indeed holds, the scaling of the energy spectra might change within this window. We do not discuss this further here.

  \section{Summary and outlook}\label{summ}
  In this work we have developed a scaling theory for fully developed incompressible hydrodynamic turbulence in a rotating fluid in the inertial range. We have studied the scaling of the energy spectra in the inertial range for weak and strong rotations, i.e., for small and large Rossby number $R_o$. We argue that for wavevectors smaller than the Zeeman wavevector rotation is important, whereas in the opposite limit, rotation is unimportant. It is therefore expected that in the former regime the scaling may be different from the K41 scaling, but in the other regime K41 scaling should ensue. The scaling theory that we developed here bears this out. 
  
  Our scaling theory reveals that in the rotation dominated regime, not only the scaling itself is anisotropic (i.e., different dependence on $k_\perp$ and $k_\parallel$), the scaling of $v_\perp$ and $v_z$ are different. Even the dynamic exponents of $v_\perp$ and $v_z$ are different, indicating weak dynamic scaling. This suggests that in the limit of a large rotation, the flow fields are dominated by only some of the velocity components. However, our theory cannot conclusively predict whether $v_\perp$ or $v_z$ will be the dominant part. Numerical simulations should be useful to make further conclusions.
  
  We have throughout assumed that the kinetic energy flux is the relevant flux (in the Kolmogorov sense), neglecting the helicity flux. However, for sufficiently large rotation, there may be a window of length scales where the helicity flux is the dominant flux. In this regime, the scaling of the energy spectra should be different. This will be discussed elsewhere.
  
  \section{Acknowledgement}
  
  AB thanks the SERB, DST (India) for partial financial support through the MATRICS scheme [file no.: MTR/2020/000406].

  \appendix
  
  \section{Reduction of the kinetic energy flux in rotating turbulence: perturbation theory}
  
  
  We now calculate the kinetic energy flux $\Pi$ in the perturbation theory. We derive (\ref{flux-full}) and  calculate $\Pi$ to lowest non-trivial order in $\Omega$. We start by eliminating pressure from the Navier-Stokes equation (\ref{rot-NS}) in a rotating frame. We obtain
  \begin{eqnarray}
   &&(-i\tilde\omega + \nu k^2) v_i + 2P_{im}({\bf k})\Omega \epsilon_{mzp}v_p({\bf k},\tilde\omega) \nonumber \\&&+ i\frac{\lambda}{2} M_{ijp} ({\bf k})\sum_{{\bf q},\tilde\omega}v_j({\bf q},\tilde\omega)v_p({\bf k-q},\,\tilde\omega - \tilde\Omega) = f_i.\label{nav}
  \end{eqnarray}
  To the lowest order in $\Omega$, we expect $\Pi$ to depend on $\Omega$ quadratically, since the energy cascade should be independent of the sense of rotation around the $z$-axis, i.e., should be the same for clockwise and anticlockwise rotations. To this order in $\Omega$, it suffices to expand   (\ref{nav}) to ${\cal O}(\Omega)$ and construct an effective equation:

  \begin{eqnarray}
   &&(-i\tilde\omega+\nu k^2)v_i +\frac{i\lambda}{2}M_{ijp}({\bf k})\sum_{{\bf q},\tilde\omega} v_j({\bf q},\tilde\Omega)v_p({\bf k-q}, \tilde\omega-\tilde\Omega) \nonumber \\ &=& f_i - 2P_{im}({\bf k})\epsilon_{mzp}\Omega \frac{f_p}{-i\tilde\omega+\nu k^2}.\label{eff-eq}
  \end{eqnarray}
Clearly, the last term on the rhs of (\ref{eff-eq}), an effective noise, is the dominant noise in the hydrodynamic limit. We use (\ref{eff-eq}) to calculate the kinetic energy flux $\Pi$ to ${\cal O}(\Omega^2)$. The flux $\Pi$ follows

\begin{widetext}
\begin{eqnarray}
 \frac{d}{dt}\langle v^2\rangle &=& -\frac{i\lambda}{2} \int_{{\bf k},\tilde\omega}\langle \bigg[M_{ijp}({\bf k})\sum_{{\bf q},\tilde\omega}v_i({\bf -k},-\tilde\omega)v_j ({\bf q},\tilde\Omega)v_p({\bf k-q},\tilde\omega-\tilde\Omega)\nonumber \\ &+& M_{ijp}({\bf -k})\sum_{{\bf q},\tilde\omega}v_i({\bf k},\tilde\omega)v_j ({\bf q},\tilde\Omega)v_p({\bf -k-q},-\tilde\omega-\tilde\Omega)\bigg]\rangle.\label{flux-main}
\end{eqnarray}
\end{widetext}

Next we iterate and expand the rhs of (\ref{flux-main}) up to the one-loop order, which gives (\ref{flux-full}) above. Now, to the linear order in $\Omega$, 

\begin{widetext}

\begin{eqnarray}
 \langle v_j ({\bf q},\tilde\Omega) v_n({\bf -q},-\tilde\Omega)\rangle &=&\frac{\langle f_j({\bf q},\tilde\Omega)f_n({-\bf q},-\tilde\Omega)\rangle}{\tilde\Omega^2 + \nu^2 q^4}-2P_{js}({\bf q})\epsilon_{szp} G_0^2(q,\tilde\Omega)\Omega \frac{ \langle f_p({\bf q},\tilde\Omega) f_n({\bf -q},-\tilde\Omega) \rangle }{i\tilde\Omega + \nu q^2} \nonumber \\&-& 2P_{ns}({\bf q}) \epsilon_{szp} G_0^2(-q,-\tilde\Omega)\tilde\Omega \frac{ \langle f_p({\bf -q},-\tilde\Omega) f_j({\bf q},\tilde\Omega) \rangle }{-i\tilde\Omega + \nu q^2}.
 \end{eqnarray}
 
\end{widetext}
Substituting in (\ref{flux-full}) and evaluating, we find $\Pi(\Omega)<\Pi (\Omega=0)$, indicating reduction of the kinetic energy flux by rotation. Since the helicity generated by the rotation scales with $\Omega$, we conclude that with a rising helicity, the kinetic energy flux is suppressed.

  \end{document}